\documentclass[aps,amsmath,amssymb,reprint,superscriptaddress]{revtex4-1}
\usepackage{dcolumn}% Align table columns on decimal point
\usepackage{bm}% bold math
\usepackage{amsmath}
\usepackage{amssymb}
\usepackage{amsfonts}
\usepackage{subfigure}
\usepackage[pdftex]{graphicx}
\usepackage[top=0.75in, bottom=0.75in, left=0.75in, right=0.75in]{geometry}
\newcommand{\beginsupplement}{%
        \setcounter{table}{0}
        \renewcommand{\thetable}{S\arabic{table}}%
        \setcounter{figure}{0}
        \renewcommand{\thefigure}{S\arabic{figure}}%
     }

\begin{document}

\title{\textbf{Measuring the repertoire of age-related behavioral changes in \emph{Drosophila melanogaster}}}

\author{Katherine E. Overman}
\affiliation{Department of Physics, Emory University}
\author{Daniel M. Choi}
\affiliation{Department of Molecular Biology, Princeton University}
\author{Kawai Leung}
\affiliation{Department of Physics, Emory University}
\author{Joshua W. Shaevitz}
\affiliation{Department of Physics and Lewis-Sigler Institute for Integrative Genomics, Princeton University}
\author{Gordon J. Berman}\email{gordon.berman@emory.edu}
\affiliation{Department of Physics, Emory University}
\affiliation{Department of Biology, Emory University}
\date{ \today} 

\begin{abstract}
Aging affects almost all aspects of an organism -- its morphology, its physiology, its behavior. Isolating which biological mechanisms are regulating these changes, however, has proven difficult, potentially due to our inability to characterize the full repertoire of an animal's behavior across the lifespan. Using data from fruit flies (\textit{D. melanogaster}) we measure the full repertoire of behaviors as a function of age. We observe a sexually dimorphic pattern of changes in the behavioral repertoire during aging.  Although the stereotypy of the behaviors and the complexity of the repertoire overall remains relatively unchanged, we find evidence that the observed alterations in behavior can be explained by changing the fly's overall energy budget, suggesting potential connections between metabolism, aging, and behavior.
\end{abstract}

\maketitle

\section*{Introduction}
Aging is a biological process that affects nearly all organisms, resulting in profound changes to their morphology, physiology, and behavior \cite{arking2006biology,ridgel2005insights,seidler2010motor}.  While there exists variability in the precise form and timing of these alterations, stereotyped patterns of aging-related change are commonly observed at scales ranging from molecules to tissues to the entire organism \cite{bishop2010neural}. However, we lack a comprehensive framework for predicting how the multifarious age-related changes at the molecular and neuronal level directly lead to behavioral changes. 

 While many age-related changes in behavior are due to direct reductions in an animal's capacity for movement (e.g., arthritis in humans or wing damage in flies), another commonly posited hypothesis is that aging effects in behavior can be partially understood as an alteration in an animal's energy budget \cite{kirkwood2000we,manini2010energy}.  In other words, while the organism may still be able to physically perform most activities within its repertoire, its reduced metabolic efficiency might impose constraints on an animal's total amount of energy to expend, leading to age-related changes in its behavioral repertoire.  This idea, that the available energy an animal possesses would have systemic effects on its chosen actions, is reminiscent of the ``hydraulic" theory of action selection that was popularized by Lorenz and others \cite{lorenz1950comparative} and might be related to molecular models of metabolic decline such as insulin pathway modifications \cite{kaletsky2010role,Akintola.2015,murphy2018insulin}.

Testing the hypothesis that age-related alterations can be understood through alterations in energy budgets, however, has proven difficult, partially due to the limitations in our ability to accurately measure full repertoires of behavior across time. Aging is a complex, dynamical process that cannot be measured at a single time-point, but, rather, it must be characterized as a trajectory across a lifetime.  Accordingly, to measure how animals' behavioral repertoires and their usage alter with age, we need to have not only a framework to measure repertoires at the timescale of single stereotyped movement (order of tens of milliseconds to seconds), but also new analysis methods to isolate the between-age-group variability from the within-age-group variability in these behaviors, finding combinations of behaviors that best describe the dynamics of aging.

In this paper, we study the age-related changes in the behavioral dynamics of the fruit fly \textit{Drosophila melanogaster}, a common model system for the study of aging and behavior \cite{le1983patterns,le1987rate,Fernandez.1999,Privalova.2021}. We measure the full repertoire of behaviors that flies of varying ages perform. While previous research on aging and behavior in flies focus on how only a small number of behaviors change with age, here, by quantifying the full repertoire of behaviors that the animals exhibit in our experimental conditions, we can observe how behavioral performance, in terms of both usage frequency and context-dependent usage (e.g., transition probabilities), changes with age. To measure the animals' behavior, we use an unsupervised method that identifies the stereotyped behaviors that the fly performs without a priori behavioral definitions - behavioral mapping \cite{berman2014mapping}.  Our results show that (1) large changes and a sexual dimorphism in how the behavioral repertoire changes with age; (2) despite these changes, the overall complexity of the flies' behavior remains unchanged; (3) as the fruit flies age, their behavioral repertoires alter, but the behaviors are still performed with similar stereotypy; (4) we can explain most of the inter-age-group behavioral variability that we observe by using an estimation of average power consumption. Thus, we provide evidence that the energy budget that an animal has available may be a key factor in regulating its behavior with age. This result encourages further investigation into the physiological basis of aging, lending credence to hypotheses that link metabolic decline to age-related behavioral changes in animals.

\clearpage

\begin{figure*}
  \centering
  \includegraphics[width=1.5\columnwidth]{spaces.png}
  \caption{Behavioral densities for quantifying the full repertoire of behaviors of male and female fruit flies (\textit{D. melanogaster}) with ages ranging from 0-70 days. (\textbf{A}) A behavioral density averaged across all flies in this study (both males and females). The color scale corresponds to the probability density function, where red peaks correspond to individual stereotyped behaviors. (\textbf{B}) Applying a watershed transform on the PDF from (\textbf{A}) produces boundary lines for the different behavioral states. Similar types of movement (described via manual annotation of the videos) are clustered together on the behavioral space, and broad descriptions of the type of movements in each cluster are obtained from the original videos. By taking the embedded points and sorting them by sex, we can make behavioral densities for the males and females separately. (\textbf{C}) The behavioral density averaged over all the male flies from all age groups. (\textbf{D}) Same as in (C), but averaged over all female flies.}
  \label{fig:maps}
\end{figure*}

\section*{Results}

\subsection*{Experiments and behavioral densities}
In order to characterize how flies' behavioral repertoires changes with age, we imaged flies (\emph{Drosophila melanogaster}) in a largely featureless environment (see Materials and Methods for details).  In total, we imaged 304 flies (155 male and 159 female), each aged between 0 and 70 days old.  To measure the flies' behavioral repertoires, we use the behavioral mapping approach originally described in Berman (2014) \cite{berman2014mapping}.  In brief, this method uses image compression techniques to measure a time series of the fly's postural dynamics, computes a continuous wavelet transform to isolate the dynamical properties of these time series (i.e., finding which parts of the body are moving at what speeds), and uses t-Distributed Stochastic Neighbor Embedding (t-SNE) to perform dimensionality reduction on the amplitudes of this transform, creating a 2-dimensional probability density function over the space of postural dynamics.  We refer to the arrangement of peaks within this probability density function as our behavioral space.  

Each peak within this density represents a distinct stereotyped behavior (e.g., grooming, running, idle, etc.).  Thus, the relative probabilities of observing a fly within each peak in the density is a measure of the animal's behavioral repertoire, seen in Figure \ref{fig:maps}A. Following the procedure  described in Cande (2018)\cite{cande2018optogenetic}, all flies - including all males and all females of all ages - were embedded into the same space in order to facilitate comparisons between individuals, sexes, and ages.  We isolate the individual peaks by applying a watershed transform \cite{meyer1994topographic} to segment the density into 122 discrete states, with near-by regions corresponding to similar behaviors (Figure \ref{fig:maps}B). The density for all the males can be seen in Figure \ref{fig:maps}C, and the density for all the females in Figure \ref{fig:maps}D. These behavioral densities provide the foundation for our analysis, as we use them to quantify how behavioral repertoires change with age.

\begin{figure*}
  \centering
  \includegraphics[width=\textwidth]{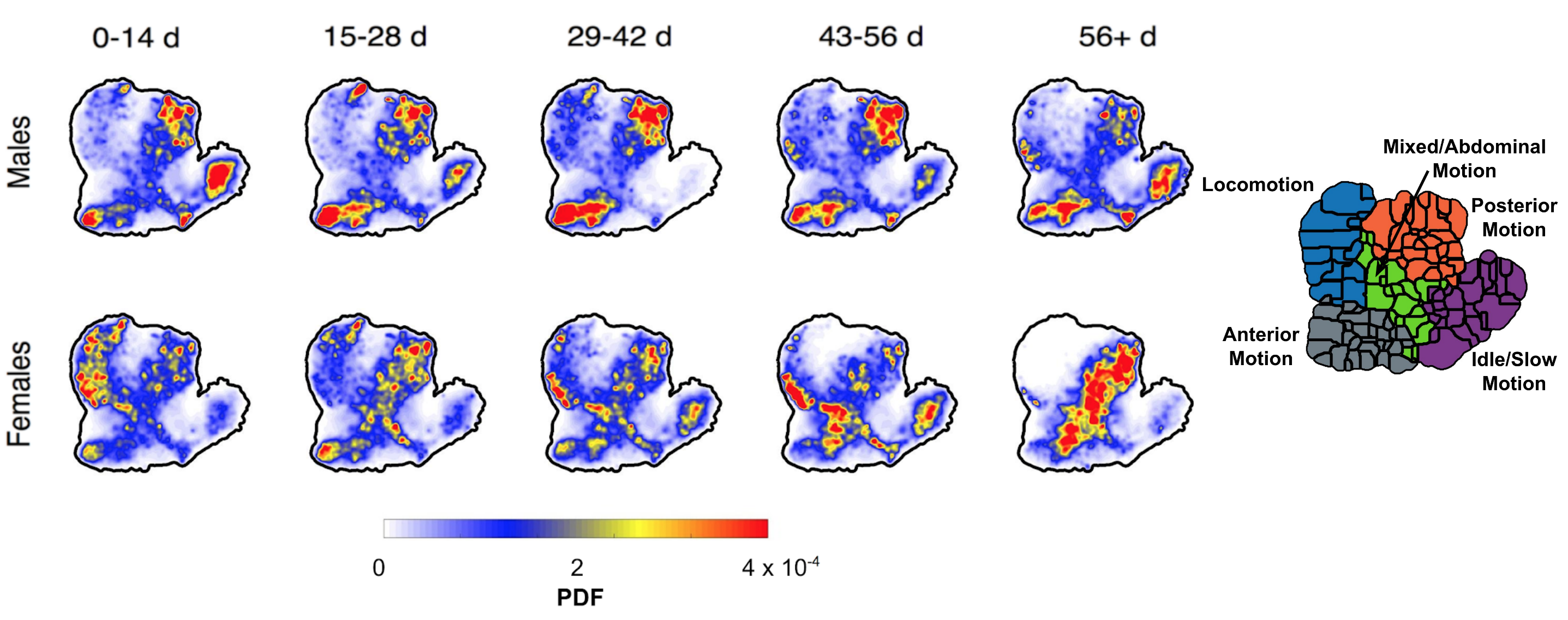}
  \caption{Behavioral densities as a function of age. (\textbf{A}) Behavioral densities for male and female flies with ages broken down into 2-week intervals. We construct these densities by separating the embedded points into subgroups of male, female, and their age. Then, we take each set of points and apply a probability density function. In this figure, we can see a broad description for behavior as a function of age emerge. Male flies mostly perform idle or slow throughout their life with the exception of mid-life, when they do more active behaviors. In contrast, females are very active when young and become more idle as they age.(\textbf{B}) Annotated behavioral space from Figure \ref{fig:maps}}
  \label{fig:ageMaps}
\end{figure*} 

\subsection*{Quantifying behavioral changes with age}

Dividing the males and females each into two-week-interval age groups (Figure \ref{fig:ageMaps}), we observe a sexual dimorphism in how their behaviors change with age. Specifically, the younger male flies mostly perform idle behaviors. In mid-life, they perform more active behaviors before again becoming lethargic in later life. Conversely, the females perform active behaviors when young, and gradually begin to perform more idle behaviors as they increase in age (excepting the last age group, which is likely under-sampled). While these results could have been found with center-of-mass tracking or other less computationally intensive methods than behavioral mapping, that our method replicates previously observed experimental results \cite{le1987rate}, provides additional confidence in the analyses to follow.

\begin{figure*}
  \centering
  \includegraphics[width=\textwidth]{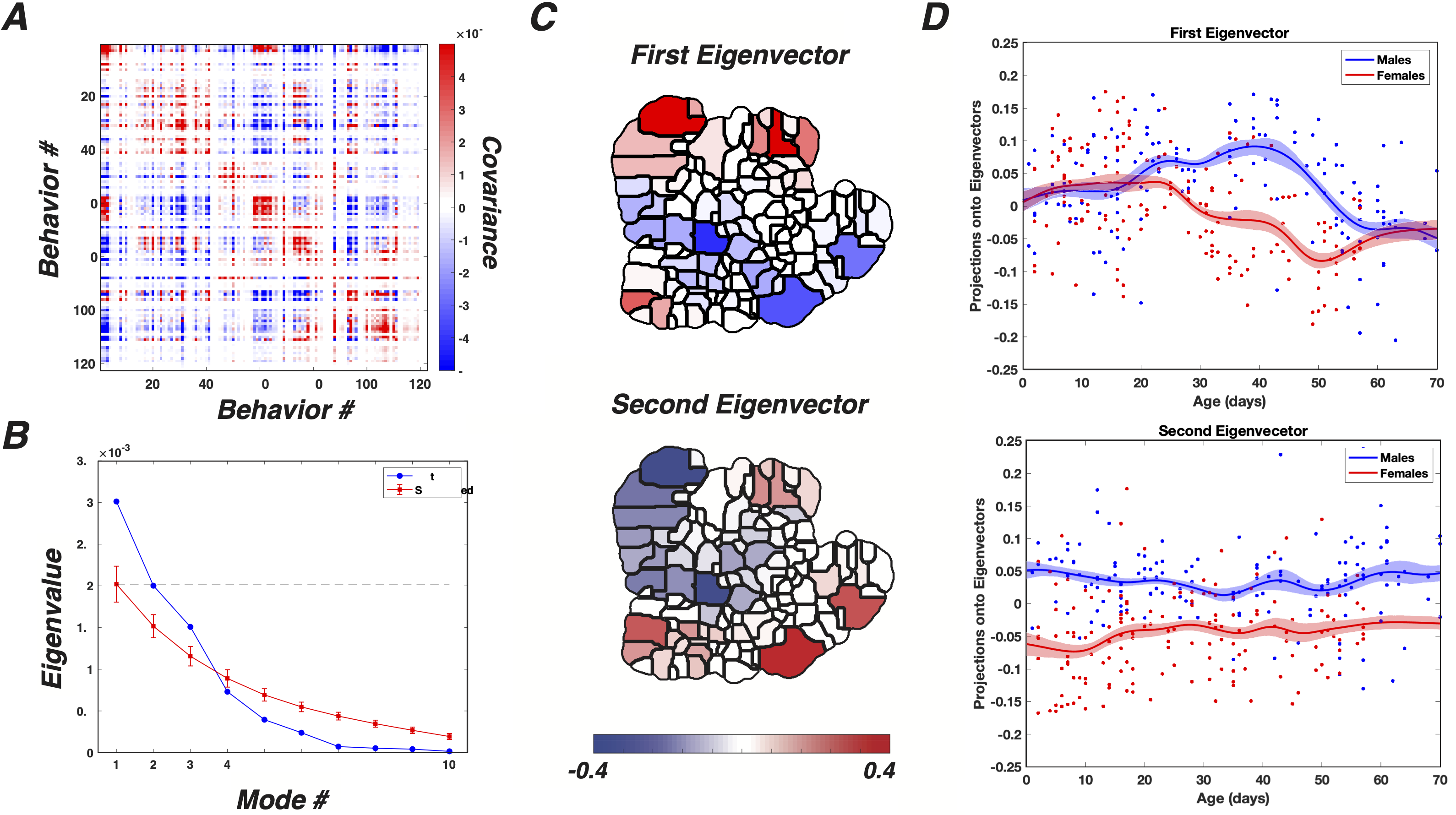}
  \caption{Identifying aging-specific behavioral covariances. (\textbf{A}) The covariance matrix of the mean behaviors sorted according to the clusters in in Figure \ref{fig:maps}B. (\textbf{B}) The eigenvalues of the covariance matrix. There are two eigenvalues (blue) that are larger or approximately equal to the eigenvalues returned from shuffling the behavioral density matrix (red, error bars are the standard deviations from many independent shuffles of the data). These two modes account for approximately 62\% of the variation in the data. (\textbf{C}) The eigenvectors corresponding to the largest (top) and second-largest (bottom) eigenvalues. (\textbf{D}) Projections of the data onto the largest (top) and second-largest (bottom) eigenvectors in (C), plotted as a function of age. Here, dots are values for individual animals, and the solid lines are from smoothing the data with a Gaussian of $\sigma=3.5$.  Error bars are the standard deviations of this process as a function of age after re-calculating the curve with re-sampled data (drawn with replacement from the original data).}
  \label{fig:cov}
\end{figure*}

While the data plotted in Figure \ref{fig:ageMaps} displays how flies' mean behavioral profile alters with age, there also exist significant variance and co-variance within sex and age groups \cite{hernandez2020framework,honegger2018stochasticity,ayroles2015behavioral}.  Thus, we need to isolate the variance in our data that is associated with changing age, rather than from inter-age-group variability.  To quantify the inter-group behavioral variance structure, we measured the behavioral covariance matrix across all sex/ages, providing a quantification of the behaviors that are shifting together with age, and the latter quantifies the within-group variability.  

Our analyses here use the discretized version of the behavioral densities, using the watershed-transform-derived regions shown in Figure \ref{fig:maps}B. $\pmb{P}^{(i)}$ is a vector of probabilities, where, $P^{(i)}_j$ is the the time-averaged probability that fly $i$ performs behavior $j$ during the one hour filming epoch -- we call this vector our behavioral vector. Given these values, we can then calculate the average  behavioral density for all individuals within each sex/age group.  We define this group-specific mean behavioral vector  to be $\pmb{\mu}^{(z)}_k$, where $z\in\{\mbox{male},\mbox{female}\}$ and $k$ is the age group.  From these means, we can then compute the covariance matrix of the set of mean behavioral vectors, $M \equiv \left[\pmb{\mu}^{(\mbox{male})}_1 \cdots \pmb{\mu}^{(\mbox{male})}_5 \pmb{\mu}^{(\mbox{female})}_1 \cdots \pmb{\mu}^{(\mbox{female})}_5\right]\in \mathcal{R}^{122\times 10}$ (5 different 2-week groups for each sex).

This covariance matrix ($C^{(M)} \equiv$ Cov($M$)), shown in Figure \ref{fig:cov}A, quantifies which behaviors are likely to increase or decrease with respect to each other across sex/age groups. To further quantify the structure within $C^{(M)}$, we calculate its eigenvectors and eigenvalues (Figure \ref{fig:cov}B-C).  Because the covariance matrix is, by definition, real-valued and symmetric, all of its eigenvalues must be greater or equal to zero.  We focus here on only the modes corresponding to the two largest eigenvalues, as only these two modes have eigenvalues that are significantly larger or similar in value to those from a covariance matrix derived from independently shuffling each of the columns in $M$.  Although there is not a clear interpretation of these two eigenvectors ($\hat{v}_1$ and $\hat{v}_2$), both appear to capture the relative performance of idle and locomotory behaviors, and the first also appears to capture the relative usage of slow vs. fast locomotion.  By plotting the projection of each fly's behavioral vector as a function of age and finding Gaussian-smoothed average curves (see Materials and Methods), we see how this low-dimensional space of behaviors alters as the flies age (Figure \ref{fig:cov}D).  There is a clear sexual dimorphism in the projections onto the first eigenmode, with the male flies exhibiting non-monotonic dynamics with age, whereas the female's average curve is largely monotonically decreasing. A similar dynamic can be observed in the second eigenmode but with a more subtle shift, as well as a sign flip. These results agree with the visual intuition from Figure \ref{fig:ageMaps} and provide a quantification of the most important changes in the flies' behavioral repertoire with age.

\subsection*{Estimated Energy Consumption Alters with Age}
As stated in the introduction, a potential mechanism for the flies' observed changes in behavior could be an overall reduction in the flies' energy budget with age.  While it was not possible to directly measure the power consumption from the animals in our experiments, we can instead estimate the metabolic cost of the observed behaviors with a biomechanical model.  

Given that the flies are constrained to move within a two-dimensional environment, we focus our modeling efforts on estimating the cost of legged locomotion within the arena (making the assumption that non-locomotion behaviors like grooming are negligible in energetic cost compared to locomotion, see Materials and Methods for further justification).  Our model of the power consumption during locomotion largely follows that of \emph{Nishi} (2006) \cite{nishii2006analytical}, which estimates the heat dissipation and work done during each swing and stance phase of locomotion at a given velocity using a biomechanical model of force production during legged locomotion (see Materials and Methods for details).  While this model has several free parameters related to the fly morphology and how gait dynamics alter with speed, we use morphological and scaling data from the literature on legged locomotion \cite{mendes2013quantification,Isakov.2016} to set these parameters. More precisely, we wish to calculate $R(v)$, the specific power (mechanical power per unit mass) required for the fly to move at a speed $v$.

\begin{figure*}
  \centering
  \includegraphics[width=.8\textwidth]{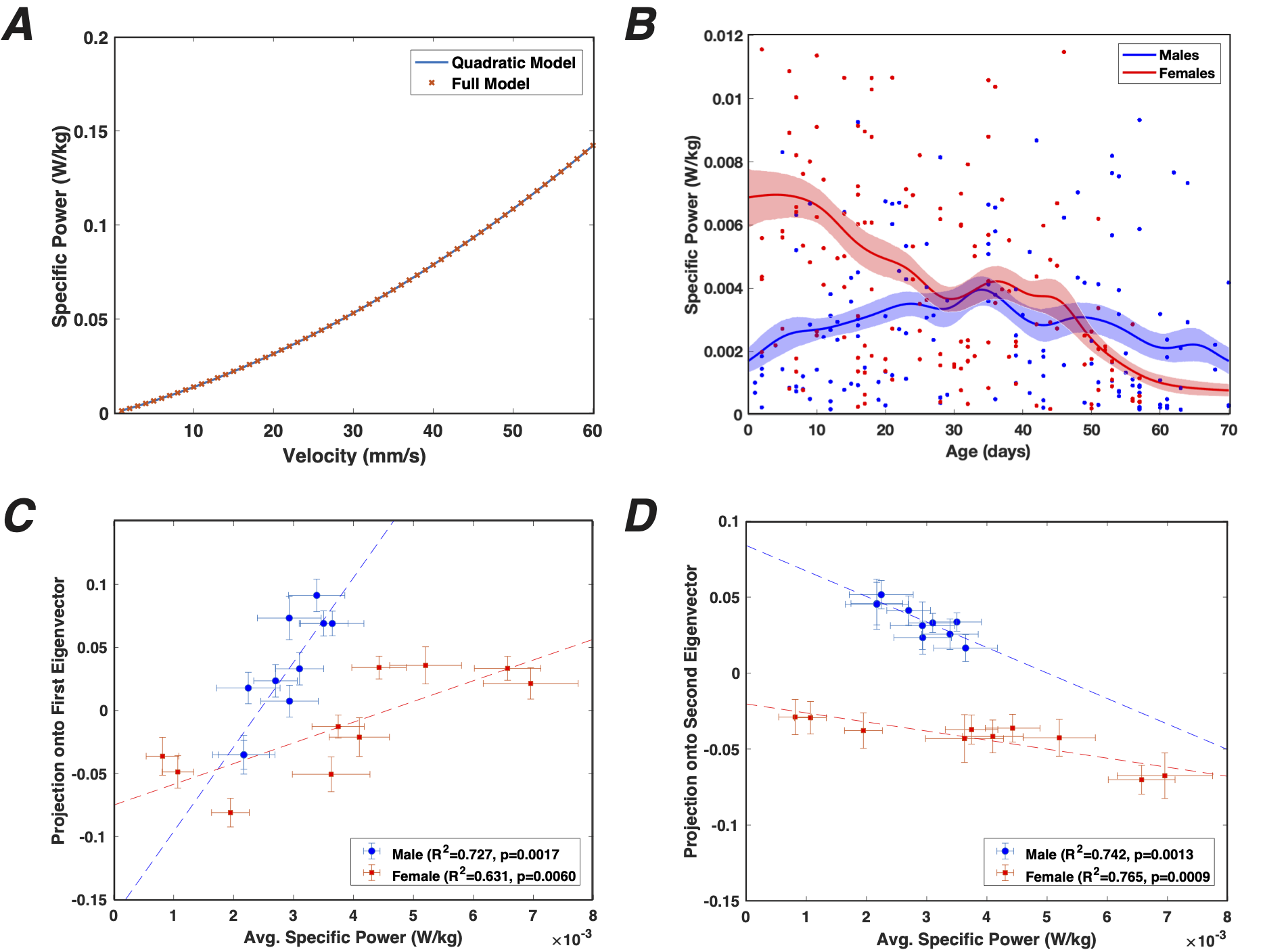}
  \caption{Energy usage predicts aging-specific changes in behavior. (\textbf{A}) Comparison of the quadratic model to the full model of Nishii (2006) to estimate specific power (power per unit mass) of legged locomotion in fruit flies. (\textbf{B}) Specific power as a function of age for male (blue) and female (orange) fruit flies. Each point represents an individual, and the curves are the Gaussian-smoothed means ($\sigma$ = 3.5 days), with error bars generated in the same manner as Fig. \ref{fig:cov}D. (\textbf{C}) Average projections onto the first eigenmode (Fig. \ref{fig:cov}C (top)) plotted versus the average specific power consumption for both male and female flies. Each point represents the value (plus error bars) from the curves in (B) and Fig. \ref{fig:cov}D (top), each spaced 7 days apart. Dashed lines are the linear fits to the data.  (\textbf{D}) Same as (C), but instead using projections onto the second eigenmode (Fig. \ref{fig:cov}C (bottom)). Note that at over 70\% of the mean aging-specific variation can be explained using the first two eigenmodes.}   \label{fig:energetics}
\end{figure*}

From tracking the center-of-mass of each fly, we are able to measure $p_i(v)$, the probability density function for speed for fly $i$, for each animal.  Given this distribution and our expression for $R(v)$, we can calculate the average specific power consumption, $\bar{R_i}$ for each animal through numerically integrating
\begin{equation}
\bar{R_i} = \int_0^{v_{max}} p_i(v)R(v) dv,
\end{equation}
where $v_{max}$ is the largest observed speed for the flies.  To make this calculation more tractable, we find that for biologically realistic range of locomotion speeds (0-60 mm/s), $R(v)$ is well-approximated by a quadratic function ($R(v) = av^2+bv+c$, where $a = 19.9 s^{-1}$, $b = 1.17 m/s^2$, and $c=.0002m^2/s^3$), as shown in Figure \ref{fig:energetics}A.
 
 The results of this calculation for each individual animal are shown in Figure \ref{fig:energetics}B as a function of age. While there is significant scatter in the data (likely due to variance in the internal activity state of the flies \cite{berman2016predictability,hernandez2020framework}), when we compute a smoothed average of the data, a clearer portrait emerges. Specifically, we observe that these curves are reminiscent of the sexual dimorphism we observed in the inter-group eigenvector projections in Figure \ref{fig:cov}.  More quantitatively, we see that when plotting the eigenvector projections versus the group-average specific power (Figure \ref{fig:energetics}C), we see a high degree of correlation for each of these values.  As seen in the figure,  we can explain at least 72\% of the aging-specific behavioral variation can be explained using a linear fit to the estimated specific power consumption. Thus, these analyses imply that most of the age-related changes we observe in the animal's behavior are correlated with changes in the average energy expenditures of the flies.

\subsection*{Complexity of the Behavioral Repertoire}
Although we show that most age-related changes in fly behavior are correlated with energy consumption, it still may be possible that other factors such as the complexity of the behavioral repertoire or the degradation of stereotyped behaviors might also be observed as the animals age \cite{Gauvrit.2017,Halberda.2012}. We test the former of these hypotheses by calculating the entropy of the behavioral space, using this metric as a proxy for the overall repertoire complexity.

\begin{figure*}
  \centering
  \includegraphics[width=.8\textwidth]{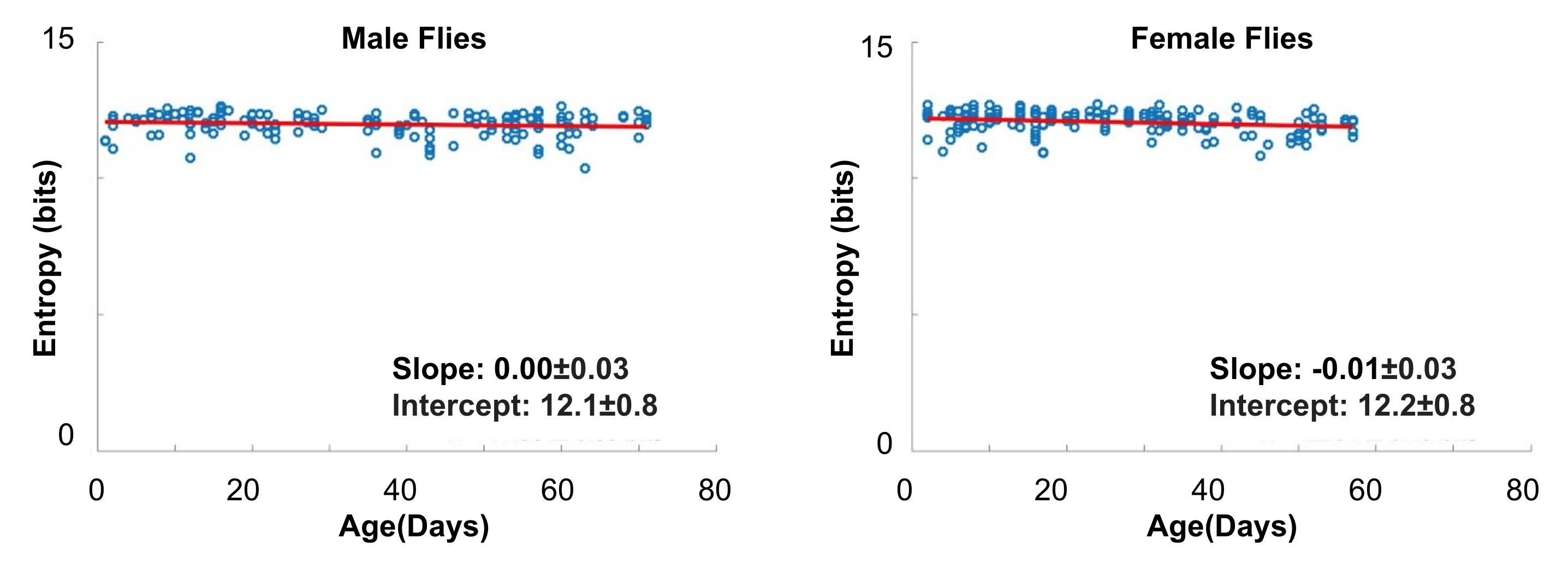}
  \caption{Entropy of the behavioral densities as a function of age for the males (left) and females (right) with a best fit line to estimate the value included on the plot by taking the mean value. The males have a slope of $-0.00 \pm 0.03$ and the females have a slope of $-0.01 \pm 0.03$.}
  \label{fig:entropy}
\end{figure*}

Specifically, we measure the entropy, $H_i$, of each individual fly's behavioral density according to 
\begin{equation}
    H_i =-\iint \! \rho(x,y) log_2\rho(x,y) dxdy,
\end{equation}
where $\rho(x,y)$ is the probability distribution over the two-dimensional behavioral space. Plotting $H_i$ as a function of the flies' ages (Figure \ref{fig:entropy}), we see no discernible trend in entropy vs. age, with the best fit slopes showing a value of $-0.00 \pm 0.03$ for the male flies and $-0.01 \pm 0.03$ for the female flies.  Thus, even though the behavioral densities are dramatically changing with age, the overall complexity remains largely unaltered, and thus we cannot conclude that the complexity of the repertoire degrades with age.

%This plot shows little change in entropy with [slopes of the best fit line here] and $H=11.9764\pm0.6336\:J/K$ for the males and $H=12.0389\pm0.4583\:J/K$ for the females, on average. In other words, even though the behavioral maps are visibly changing with age, overall their complexity neither increases or decreases implying the underlying dynamics should be constant as well.

\subsection*{Long Time Scales and Hierarchical Structure in Behavior with Age}
While the complexity of the behavioral repertoire remains unchanged, the complexity of how the animals traverse through this space over time might still show significant deviations. Prior investigations into the complexity of fly behavioral sequences have shown that these dynamics of transitions between stereotyped behaviors exhibit long time scales and hierarchical organization \cite{berman2016predictability,hernandez2020framework}.  A hypothesis for aging-related behavioral change is that the structure of the behavioral repertoire becomes less complex with age \cite{Gauvrit.2017,Halberda.2012}, and with the detailed measurements of behavior described here, we can test this idea, potentially gaining insight into changes occurring to the internal programs that may generate these patterns. 

\begin{figure*}
  \centering
  \includegraphics[width=.8\textwidth]{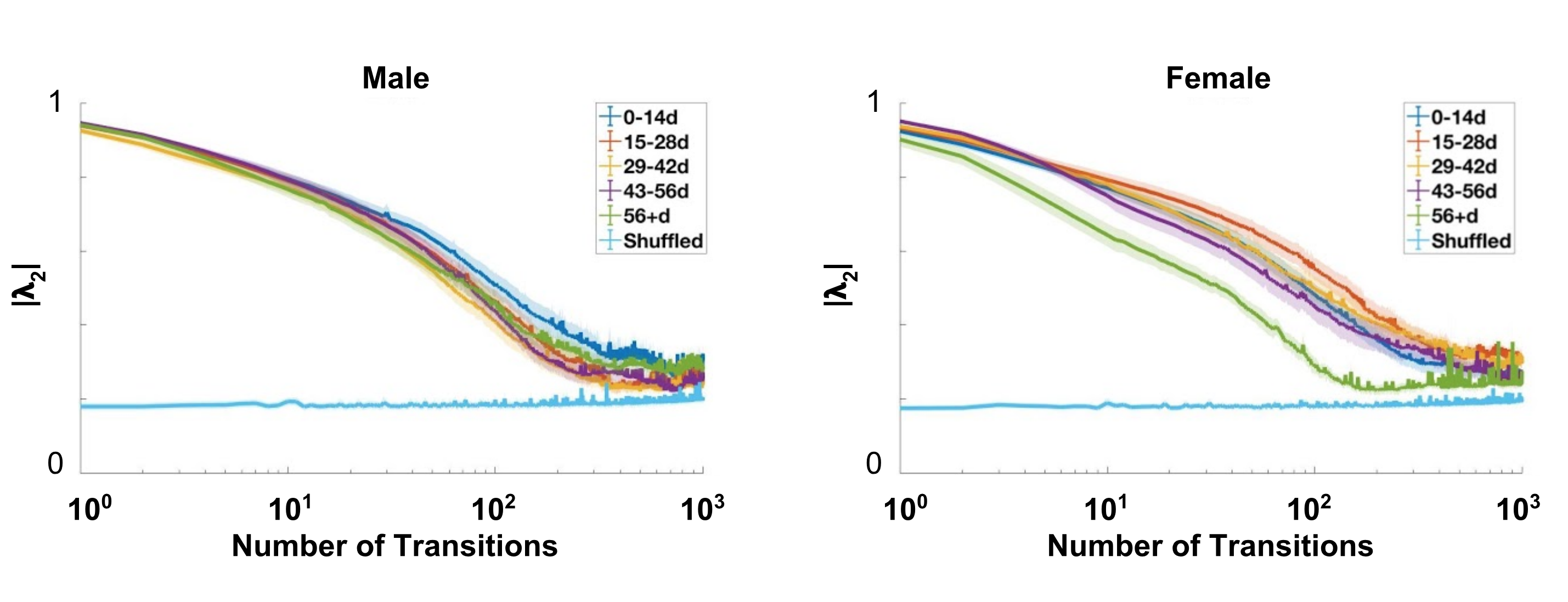}
  \caption{Absolute value of the second eigenvalue of the transition matrices as a function of transitions into the future, averaged over all flies in each age group with error bars corresponding to standard deviation of the mean for the male flies (left) and the female flies (right). The light blue line, which acts as a noise floor, is the second eigenvalue in a transition matrix calculated after shuffling our finite data set.}
  \label{fig:TimeScales}
\end{figure*}

First, to assess the overall timescale structure of the flies' behavioral patterns, we measure the transition matrix at different time scales, 
\begin{equation}
    [\mathbf{T}(\tau)]_{i,j}=p(S(n+\tau)=i|S(n)=j),
\end{equation}
 where $i$ and $j$ as two stereotyped behaviors, $S(n)$ is the behavioral state of a system at transition $n$ (note: to decouple waiting time in a state from complexity in the order of pattern of transitions between states, we measure time in units of transitions, following the methods in \cite{berman2016predictability}).  We can decompose each of these matrices via
 \begin{equation}
     [\mathbf{T}(\tau)]_{i,j}=\sum_\mu\lambda_\mu(\tau)u_i^\mu(\tau)v_j^\mu(\tau),
 \end{equation}
where $u_i^\mu$ and $v_j^\mu$ are the $i^\text{th}$ right and $j^\text{th}$ left eigenvectors of the matrix, respectively, and $\lambda_\mu$ is the eigenvalue with the $\mu^\text{th}$ largest modulus. Because the columns of each of these matrices must sum to one, $\lambda_{1}(\tau)=1$ for all values of $\tau$, and $|\lambda_{\mu>1}(\tau)|<1$ by the Perron-Frobenius Theorem. While for a Markov Model, the eigenvalues should decay exponentially with $\tau$, we find that flies in all sex and age groups exhibit super-Markovian time scales (Figure \ref{fig:TimeScales} shows the results for the second-largest eigenvalues in each transition matrix).  With the exception of the $>56$ day-old females (for which we had fewer individuals in our sample), however, we found no significant difference between the time scales across age groups.

%It is known that these eigenvalues of the behavioral transition matrix produce long, non-Markovian time scales describing the loss of predictability over time \cite{berman2016predictability}. We measure all $\mu=122$ of the times scales for each age group keeping males and females separate. For visibility and since it has the largest magnitude and the most information about predictability, we have shown only $\lambda_2$ in Figure \ref{fig:TimeScales} with the male age groups on the left and the female age groups on the right. With the exception of the 56+d females, which has less data (and therefore less information about the future by default) than the other age groups, we see no significant difference between the time scales across age groups. Therefore, the predictability of the fruit flies doesn't change with age.

While the complexity of the repertoire or the overall timescale might not be changing with age, the underlying structure of the behavioral transitions might still be altering.  To test for this possibility, we applied a predictive clustering analysis to the space to identify groupings of behaviors that best preserve information about the long timescale structure in our data.  More precisely, we would like to find a partition of our behavioral space, $Z$, such that this representation has a simple of a representation as possible, while still maintaining information about the future behavioral states of the animal.  Here, we achieve this using the Deterministic Information Bottleneck (DIB) approach  \cite{bialek2001predictability,strouse2017deterministic}, which minimizes the functional
\begin{equation}
    L_{DIB}=H(Z)-\beta I(Z;S(n+\tau)),
\end{equation}
where $Z$ is our partition, $H(Z)$ is the entropy of the partition, and $\beta$ is a Lagrange multiplier that modulates the relative importance of simplicity and predictability.  We perform this optimization for several values of $\tau$ for each age group, in all cases varying $\beta$ and the number of initial clusters in $Z$ to create a full curve of values (see Materials and Methods for details).  

\begin{figure*}
  \centering
  \includegraphics[width=.8\textwidth]{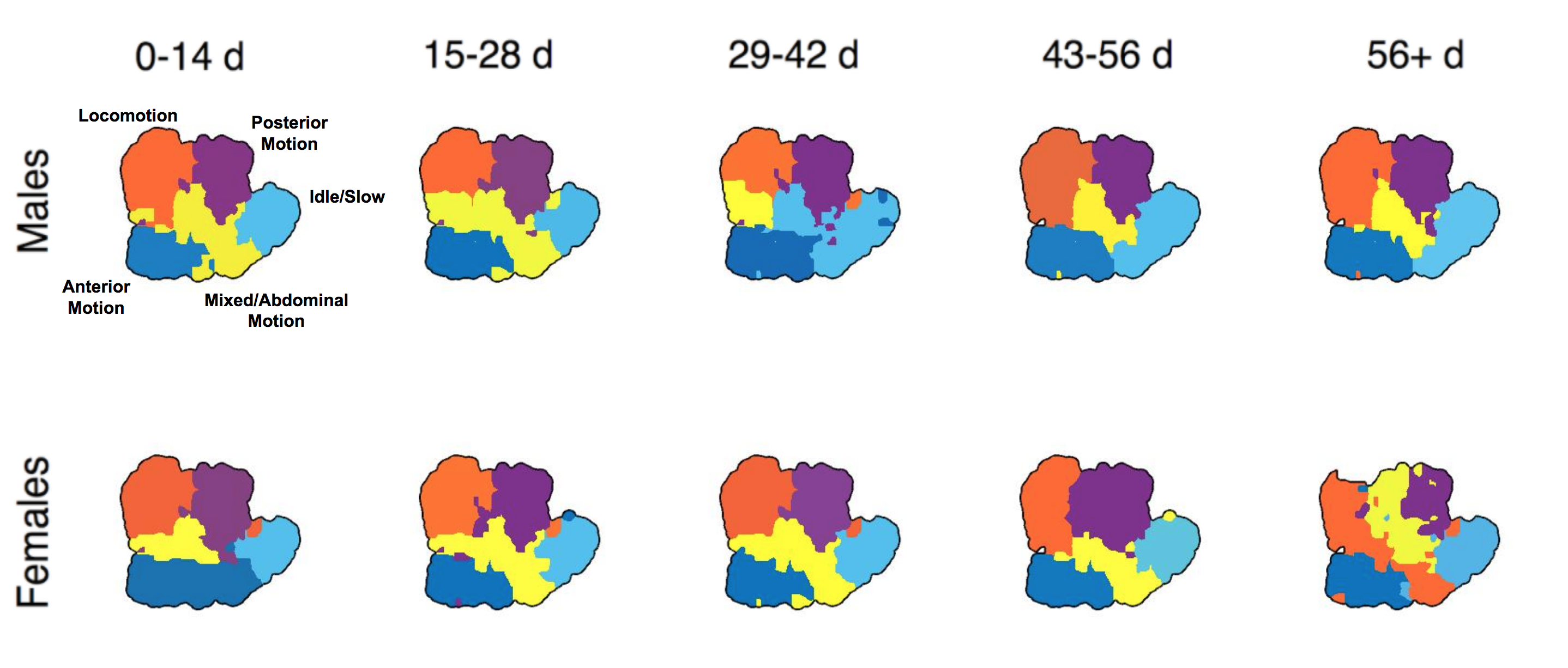}
  \caption{Hierarchical partitioning solutions from deterministic information bottleneck for the behavioral density with $\tau=100$ and 5 clusters for male flies (top) and female flies (bottom) as a function of age.}
  \label{fig:DIB}
\end{figure*}

The resulting clusterings for $\tau$=100 with five clusters can be seen in Figure \ref{fig:DIB}. As with the eigenvalues in the previous plot, the clusters obtained via this approach remain nearly constant with varying age, with only small-probability behaviors flipping between regions.  Thus, we lack evidence of significant alterations of the temporal complexity of the flies' behavior with age.  

\subsection*{Stereotypy}
Lastly, while we observe no significant changes to the flies' repertoire or temporal complexity, we still can measure if there is deterioration in how the behaviors are performed, potentially implying that the flies are undergoing a physical deterioration or some other inability to consistently perform behaviors while aging.  To assess changes in how stereotyped behaviors are performed, we measure how much the performance of individual behaviors are altered with age, quantifying a decreased stereotypy with an increase in the variance of the postural trajectories underlying the performance of these actions. 

We divide the data into age groups of two week intervals, with a one week overlap (0-14 days, 8-21 days, 15-28 days, etc.), finding the postural trajectories associated with the performance of each behavior.  While the details of this can be found in Materials and Methods, broadly, we use a phase-reconstruction method (based on Revzen (2008)\cite{revzen2008estimating}) across all of the postural modes for each time a behavior is performed.  We measure the mean postural dynamics across all individuals in a given sex/age group and assess the stereotypy of each behavior ($b$) in each age group ($\kappa$) with our Stereotypy Index, $\chi_{b,\kappa}$, which is the fraction variance explained by the mean trajectory for that behavior.  Thus $\chi_{b,\kappa}\to 1$ implies that each time the behavior is performed, its  postural trajectories are exactly the same (maximally stereotyped), and $\chi_{b,\kappa}\to 0$ implies that the postural trajectories are different each time the behavior is performed (minimally stereotyped).

\begin{figure*}
  \centering
  \includegraphics[width=.8\textwidth]{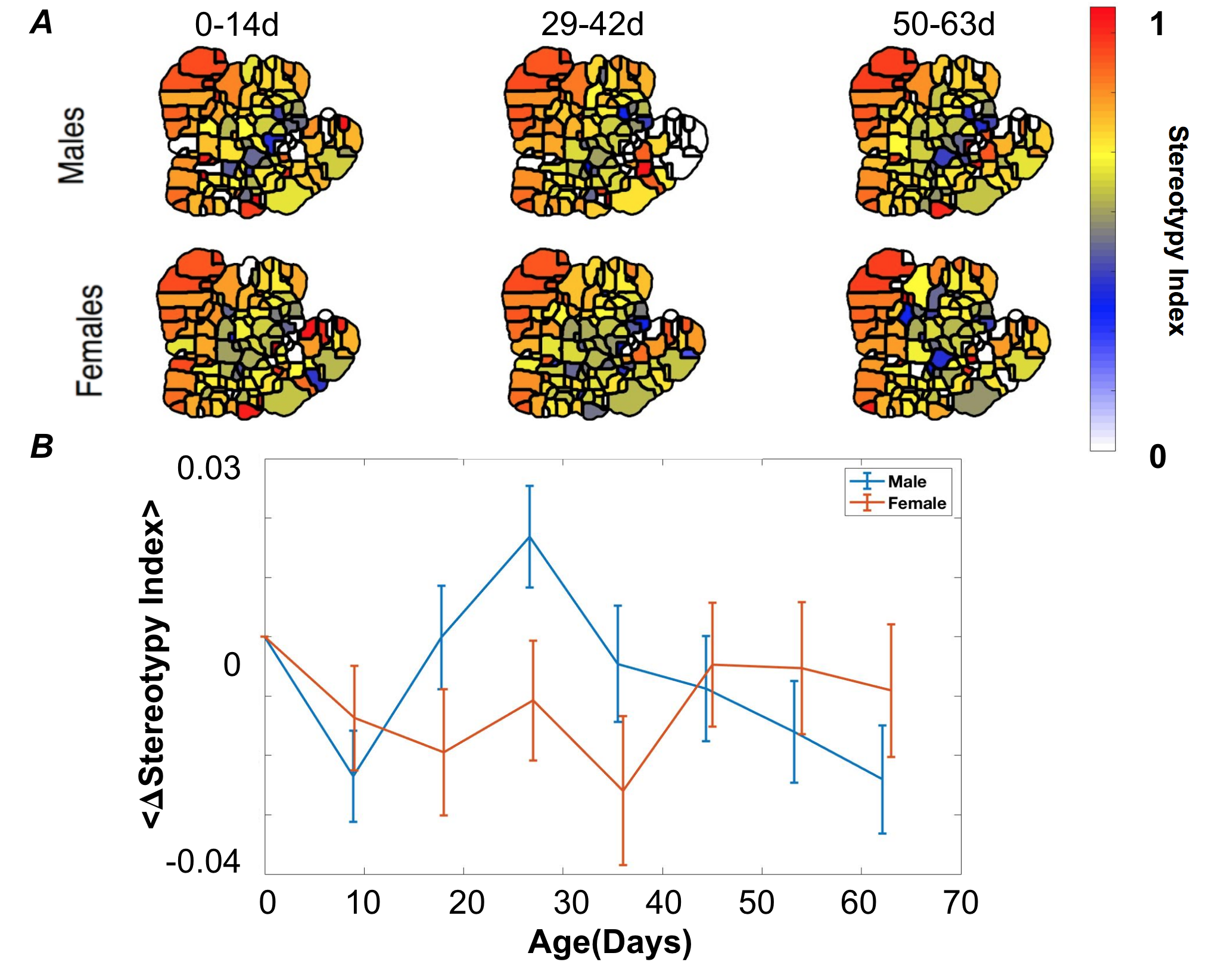}
  \caption{No significant changes in stereotypy with age. (\textbf{A})The stereotypy of each behavior (or how stereotyped each behavior is - 1 being very stereotyped, 0 being not at all stereotyped) plotted as a function of age by calculating the maximum synchronization parameters of each behavior for the males (top) and  females (bottom). (\textbf{B}) Quantification of how the synchronization parameters change between each age group and the initial age group by taking the mean difference between the synchronization parameters of each behavior.  Error bars are bootstrap estimates from re-sampling individuals with replacement.}
  \label{fig:SPs}
\end{figure*}

The values of $\chi_{b,\kappa}$ for each behavior and three different age groups are displayed in Figure \ref{fig:SPs}A. By eye, we can see only minimal changes across the age groups (and no statistically-significant changes when accounting for multiple hypothesis corrections).  Note that a few behaviors, while stereotyped, were not performed enough to get a good estimate of their synchronization parameters so those behaviors are listed as having a synchronization parameter of 0. 

To quantify this lack of change across the whole behavioral repertoire, we calculated the average stereotypy for each age group, 
\begin{equation}
\bar{\chi_\kappa} =\frac{1}{N_\kappa}\sum_b \chi_{b,\kappa} \sum_{i\in G_\kappa} P_b^{(i)},
\end{equation}
where $G_\kappa$ is the set of all flies in age group $\kappa$, $N_\kappa$ is the number of flies in the group, and $P_b^{(i)}$ is the fraction of time that fly $i$ performs behavior $b$. We then measured the difference in the average stereotypy the youngest age group and each of the subsequent age groups for each sex ($\bar{\chi_\kappa} - \bar{\chi_0}$). Figure \ref{fig:SPs}B shows the results of this calculation for both the males and females. Although we do observe some changes between the age group, they are within 1.5 standard deviations. Thus, although the probability of choosing a behavior changes with age, each behavior, when performed, is, on average, no less stereotyped.

\section*{Discussion}
In this paper, we measured the behavior of fruit flies (\textit{D. melanogaster}) at many points along their lifespan, aiming to isolate patterns of behavioral change with age and to make predictions about the physiological basis of these changes.  Consistent with previous studies, we found a sexual dimorphism in changes in the animals' overall activity level, but we also identified subtler patterns of change with age by measuring the largest eigenvalues, and their corresponding eigenvectors, of the inter-age-group covariance matrix. Despite observing no significant changes in the repertoire complexity or stereotypy with age, we find that most of the age-specific behavioral alterations can be explained by a model of energy consumption, implying that energy budget may play an overarching role in regulating aging behavior.

This observation that energy may play a key role in aging-specific changes in behavior is in accordance with results from long-lived mutants in a variety of species, many of which have changes in gene regulation pathways that affect energy availability \cite{Kenyon.2005}.  For example, mutations in the insulin/IGF-1 receptors or homologs, which promote food storage and cell replication, have been shown to extend lifespan in flies \cite{Tatar.2001,broughton2009insulin}, nematodes \cite{Kenyon.1993,murphy2018insulin}, and rodents \cite{Holzenberger.2003}.  In addition, another long-lived fly mutant, the E(z) histone methyltransferase heterozygous mutation, is associated with large alterations in a variety of metabolic regulation pathways \cite{Moskalev.2019}. In addition, these changes were found to exhibit sex-dependent effects, similar to our results as well.

In future efforts where behavioral repertoire and metabolic state could be simultaneously assayed (through, for example, proteomic or transcriptomic measurements), we would expect to find correlations between position along the the curves seen in Fig.~\ref{fig:cov}D and key metabolic regulators.  Through this methodology, it may be also possible to provide an effective age for each individual in a heterogeneously aging population, providing a phenotyping tool for identifying new molecules involved in increased and decreased longevity, as well as for the study of evolutionary aging dynamics.  

While the analysis framework detailed in this paper should be generalizable to other data sets, including other species \cite{Marshall21,Klibaite.2021} and neuroimaging data \cite{Billings.2017}, the data used in this study present several limitations that need to be studied in future work.  First, despite the wide range of behaviors we observed in our assay, many natural behaviors, including courtship and flying, were not measured here. Flight in particular is likely a large source of oxidative stress and potential injury for the animals, likely creating more opportunities for decreased stereotypy and the degradation of behavioral performance.  Additionally, due to technical constraints in our experimental set-up, we only imaged flies for one hour during their life.  Future studies would benefit from having longer recording epochs -- up to the animal's full lifetime -- that could capture the influence of circadian rhythms and could more ably measure inter- vs. intra-individual variability across the lifespan.

Despite these limitations, this study points a way forward for using full repertoires of behavior to study aging and its physiological underpinnings.   Although many of our energy budget-related analyses here could have been performed using center-of-mass tracking alone, by studying multiple actions simultaneously, it becomes not only possible to identify the age-relevant behavioral changes (here, primarily related to locomotion and slow/idle behaviors), but also to control for other possibilities such as the complexity of the animal's usage of its behavioral repertoire or behavioral degradation and to isolate covariances between and within age groups. These measurements allows us to better predict how genetic or neural manipulations may affect aging across individuals and across the lifespan, as well as to make more specific predictions as to what types of physiological factors might play a role in these changes.

\section*{Materials and methods}
\subsection*{Data}
The data consist of 304 flies (D. melanogaster), 150 of which are male and 154 of which are female, with ages ranging from 0 to 70 days of age. Each fly was imaged from above for an hour while contained in a featureless dish with sloped sides to prevent aerial movements, following the approach detailed in \cite{berman2014mapping}. Flies were placed into the arena using aspiration and provided 5 minutes to adapt to their environment before data collection. To reduce the effect of circadian rhythms, all recordings occurred between 09:00 and 13:00. The temperature was kept constant at 25$^{\circ}\pm$1$^{\circ}$C.

\subsection*{Behavioral Densities}
We created our behavioral densities following the data pipeline outlined in \cite{berman2014mapping}. This approach begins with image analysis (segmentation and alignment), projecting images onto postural eigenmodes, Morlet wavelet transforms \cite{goupillaud1984cycle}, and a dimensionally reduced embedding via t-distributed Stochastic Neighbor Embedding \cite{maaten2008visualizing}. We applied a watershed transform \cite{meyer1994topographic} to a Gaussian-smoothed density of the resulting points to isolate the individual peaks.  We defined behavioral epochs as lengths of time lasting at least 0.05s with low speeds in the behavioral densities, again following the approach of \cite{berman2014mapping}.

\subsection*{Gaussian-smoothed Average Curves}
For Figure \ref{fig:cov}D, we applied a Gaussian-smoothed average according to the following equation:

\begin{equation}\label{E:smooth}
y(t)=\frac{\sum_{i=1}^{N}e^{\frac{(t_i-t)^2}{2\sigma^2}}\cdot X_i^2}{\sum_{i=1}^{N}e^{\frac{(t_i-t)^2}{2\sigma^2}}}
\end{equation}
% \begin{equation*}
% \sigma_y(t)=\frac{\sum_{i=1}^{N}e^{\frac{(t_i-t)^2}{2\sigma^2}}\cdot\sigma_{X_i}^2}{\sum_{i=1}^{N}e^{\frac{(t_i-t)^2}{2\sigma^2}}}
% \end{equation*}
Here, $t$ is age, $X$ is the original value of the eigenvector projections, $y$ is the smoothed value of $X$, $N$ is the number of flies, and $\sigma$ corresponds to the standard deviation of the projections. For example, Figure \ref{fig:cov}D is a plot of $y$ vs. $t$.

Error bars for these plots are generated through a bootstrapping procedure.  Specifically, the data ($\{t_i,X_i\}$) are sampled with replacement, and (\ref{E:smooth}) is now applied to this re-sampled data set.  This procedure is repeated $1,000$ times (each independently sampled), and the error bars are the standard deviations of these re-sampled curves at each point in time.

 \renewcommand{\arraystretch}{1.5}
\begin{table*}
\centering
\begin{tabular}{ |c|c| } 
 \hline
 Body Weight, $M$ & $2.5\times10^{-6}$ kg \cite{Isakov.2016} \\ 
 \hline
 Body Length, $L$ & $2.5\times10^{-3}$ m \cite{Isakov.2016}  \\ 
 \hline
 Stance Length, $S$ & $(.0472\times V+ .748)/1000$ m \cite{mendes2013quantification}  \\ 
 \hline
 Velocity, $V$ & $0-6\times10^{-2}$ m/s \cite{Isakov.2016} \\ 
 \hline
 Length of Leg, $l$ & $1.3\times10^{-3}$ m \cite{Isakov.2016}\\
 \hline
 Moment of Inertia of the Leg, $I$ & $1.6\times10^{-14}$ kgm$^2$ \cite{Isakov.2016} \\
 \hline
 Duty Ratio, $\beta$ & $\frac{t_{st}}{t_{st}+t_{sw}}$ \\
\hline
 Stance Duration, $t_{st}$ & $11.5+.910V$ s \cite{mendes2013quantification} \\
\hline
 Swing Duration, $t_{sw}$ & $(-0.126V+36.56)/1000$ s \cite{mendes2013quantification} \\
 \hline
\end{tabular}
\caption{Parameters Used for Locomotion Energetics Calculations}
\label{tab:params}
\end{table*}

\subsection*{Synchronization Parameter}
By treating the fruit flies' postural modes as a phase-locked oscillator, we use the \textit{Phaser} algorithm \cite{revzen2008estimating} to estimate the behaviors' phases, providing a measure of stereotypy. For each behavior, we use the algorithm to map the individual behavioral bouts to a phase variable between 0 and $2\pi$, providing us with a phase reconstruction of our data that we can compare to the original trajectories (the methodology is the same as in \cite{berman2014mapping}). To ensure the phase-averaged orbits are aligned between individuals and bouts, we calculate the maximum cross-correlation value between orbits for every postural mode separately, which gives our phase offset. After determining which modes contribute to each behavior (we use only modes that have mode-specific synchronization parameters of greater that 0.1), we calculate the synchronization parameter for age group $\kappa$ for each behavior $b$ across all postural modes $\gamma$ according to:  
\begin{equation}
\mathcal{X}_{b,\kappa}=\frac{1}{N^{(\gamma)}_\kappa}\sum_\gamma\left[1-\frac{\sigma^2(\vec{y}{(\gamma)}_{b,\kappa}(\phi)-\vec{\mu}{(\gamma)}_{b,\kappa}(\phi))}{\sigma^2(\vec{y}{(\gamma)}_{b,\kappa}(\phi))}\right ], 
\end{equation}
where $\vec{y}{(\gamma)}_{b,\kappa}(\phi)$ contains the postural projection time series from every bout of behavior $b$, $\vec{\mu}{(\gamma)}_{b,\kappa}(\phi)$ is the phase-averaged orbits for the projection data in $\vec{y}{(\gamma)}_{b,\kappa}(\phi)$, $N^{(\gamma)}_\kappa$ is the number of postural modes used, and $\sigma^2(x)$ is the variance in $x$.

By taking the maximum value across the modes, we quantify our stereotypy for each behavior. This value ranges from 0 to 1, where 0 signifies no stereotypy and 1 signifies full stereotypy.  This algorithm requires many bouts of each behavior in order to make the calculation.

\subsection*{Deterministic Information Bottleneck}
The deterministic information bottleneck algorithm is an iterative algorithm that obeys a set of self-consistent equations:
\begin{equation}
 q(t|x) = \frac{1}{\mathcal{Z}(x,\alpha,\beta)}exp\left[\frac{1}{\alpha}(log q(t)-\beta D_{KL}[p(y|x)q(y|t)])\right]
\end{equation}
\begin{equation}
 q(t) = \sum_xp(x)q(t|x)
 \end{equation}
 \begin{equation}
 q(y|t) = \frac{1}{q(t)}\sum_xq(t|x)p(x,y)
 \end{equation}
Here, $x \in S(n)$, $y \in S(n+\tau)$, $t \in Z$, $\mathcal{Z}$ is a normalizing function, and $D_{KL}$ is the Kullback-Leibler divergence between two probability distributions. For a given $|Z|=K$ number of clusters, inverse temperature $\beta$, and random initialization of $q(t|x)$, the equations are iterated until $(\mathcal{F}_t-\mathcal{F}_{t+1})/\mathcal{F}_t<10^{-6}$ is satisfied. We performed 24 replicates of the solution using a range of $\beta \in [0.01, 500]$ spaced exponentially, $K \in [2, 30]$, and $\tau \in [1, 4096]$. The optimization is done for each value of $\beta$ until the convergence criterion is satisfied. The resulting solution is then used as the initial condition for the next value of $\beta$.

\subsection*{Power Estimation Model}
We used the model from Nishii (2006)\cite{nishii2006analytical} to estimate the power consumption according to the following equations. The swing and stance phase describes the portion of motion where the leg is sweeping forward and when the leg applies pressure to the ground to propel the body forward, respectively.

Specifically, we model the power consumption using the following equations:
\begin{align}
H^{st}&=\gamma\int_{T^{st}}(|\tau^{st}(t)|^k + |\alpha N(t)|^k)dt=\gamma\bigg(\frac{M}{n}\bigg)^2\frac{T}{\beta}\bigg(\alpha^2+\frac{S^2}{12}\bigg) \\
H^{sw}&=\gamma\int_{T^{sw}}|\tau^{sw}(t)|^kdt=\gamma\frac{2\pi^2I^2}{l^2}\frac{\beta V^3}{S(1-\beta)^3} \\
W^{st}&=\int_{T^{st}}f(N(t)x(t))\frac{V}{l}dt=\frac{MS^2}{8nl\beta}\\
W^{sw}&=\int_{T^{sw}}f(\tau^{sw}\dot{\theta})dt=I\bigg(\frac{V}{l}\bigg)^2\frac{1+\beta^2}{(1-\beta)^2}.
\end{align}
Here, $H^{st}$ is the heat dissipation during the stance phase, and $H^{sw}$ is the heat dissipation during the swing phase. Similarly, $W^{st}$ and $W^{sw}$ denote the mechanical work done during the stance and swing phase, respectively. In these equations, n is the number of legs, $\gamma$ represents the ratio of heat dissipation to mechanical work, and $\alpha$ is the amplitude of the torque required to maintain a bent leg posture. The rest of the parameters are defined in Table \ref{tab:params}. We use values to calculate the specific power as a function of velocity, which we called $e$. We calculate $e$ by summing together the power consumed from the heat and work during the stance and swing phase according to the following equations for $e^{st}_h$, $e^{sw}_h$, $e^{st}_w$, $e^{sw}_w$:
\begin{align}
e(V,\beta,S)&=e^{st}_h+e^{sw}_h+e^{st}_w+e^{sw}_w\\
e^{st}_h&=\frac{\sum_{i=1}^nH^{st}_i}{MVT}=\frac{\gamma M}{n\beta V}\bigg(\alpha^2+\frac{S^2}{12}\bigg)\\
e^{sw}_h&=\frac{\sum_{i=1}^nH^{sw}_i}{MVT}=\gamma\frac{2n\pi^2I^2}{l^2M}\frac{V^3\beta^2}{S^2(1-\beta)^3}\\
e^{st}_w&=\frac{\sum_{i=1}^nW^{st}_i}{MVT}=\frac{S}{8l}\\
e^{sw}_w&=\frac{\sum_{i=1}^nW^{sw}_i}{MVT}=\frac{nI\beta}{MS}\bigg(\frac{V}{l}\bigg)^2\frac{1+\beta^2}{(1-\beta)^2},
\end{align}
where $T$ is the gait cycle period.

Using this model, we can estimate the relative mechanical cost of grooming compared to locomotion by the quantity $\frac{e^{sw}_h+e^{sw}_w}{e^{st}_h+e^{st}_w}$, since the animal is moving its legs but is no longer having to expend excess energy to propel itself forward during the stance phase. Across all speeds, this ration is $\approx 10^{-7}$, justifying our treatment of all zero-velocity epochs as having the same energetic cost.

\section*{Acknowledgments}
K.E.O., and K.L., and G.J.B. were supported by NIMH (R01 MH115831-01), the Human Frontier Science Program (RGY0076/2018), and a Cottrell Scholar Award, a program of the Research Corporation for Science Advancement (25999).  K.E.O. was supported by the NSF Physics of Living Systems Student Research Network (PHY-1806833). Experimental work and J.W.S and D.M.C. were supported by NIH R01 GM098090 and in part by the National Science Foundation, through the Center for the Physics of Biological Function (PHY-1734030).

%\pagebreak

%\section*{Supplemental Figures}
\beginsupplement
\begin{figure*}[h!]
  \centering
  \includegraphics[width=\textwidth]{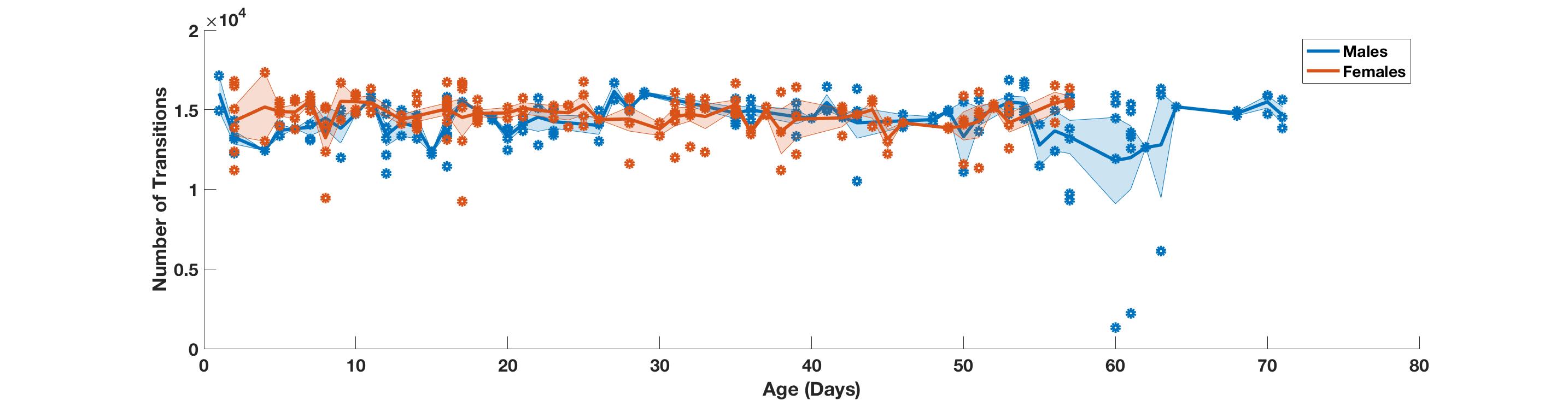}
  \caption{Average number of transitions per hour as a function of age. Each data point is a different animal, and the line is the Gaussian-weighted average (error bars are standard error of the mean for the average).}
  \label{fig:seriesLengths}
\end{figure*}

\begin{figure*}[h!]
  \centering
  \includegraphics[width=\textwidth]{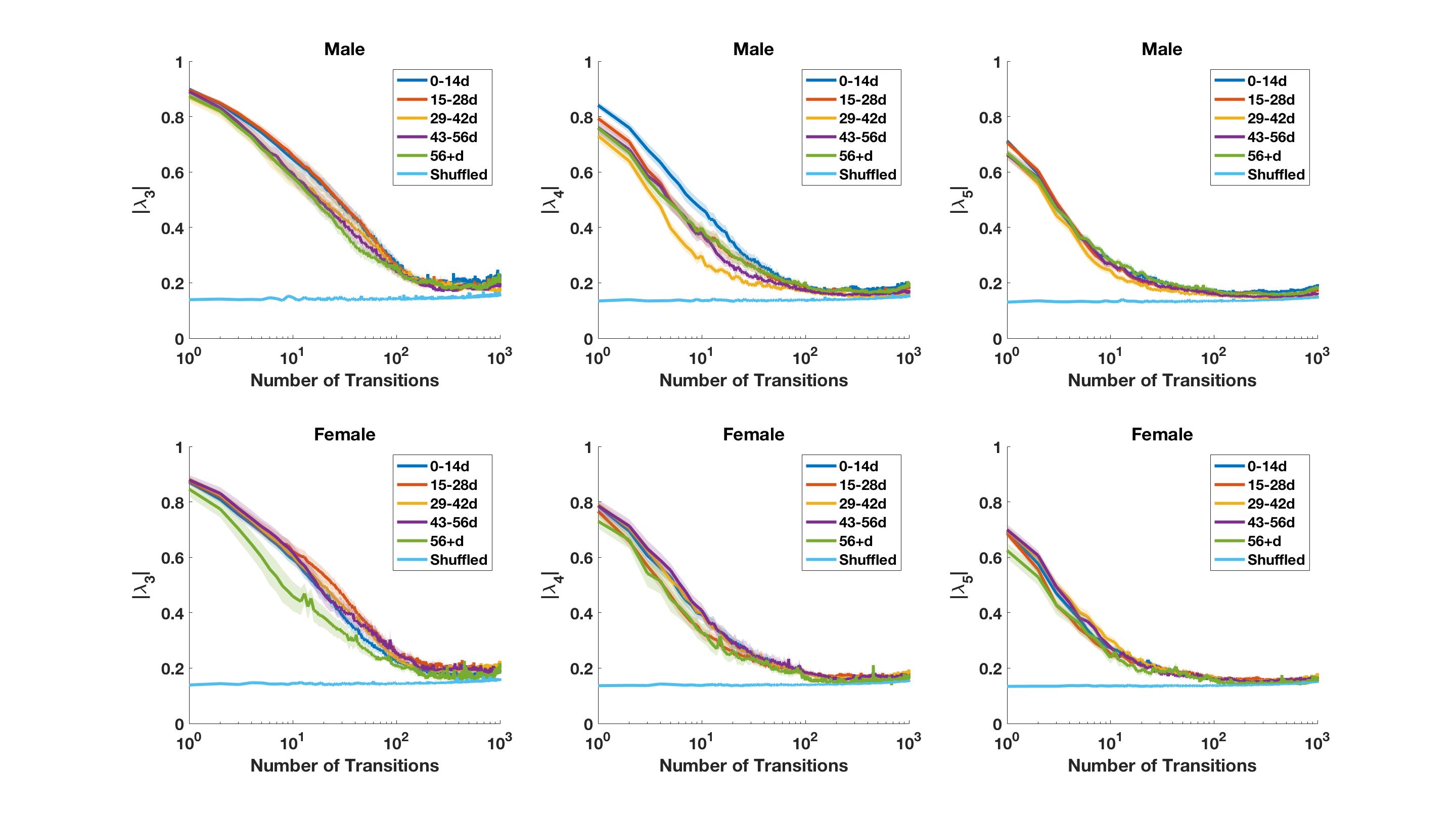}
  \caption{The third, fourth, and fifth eigenvalue timescales for each sex and age group. Line thicknesses represent the standard errors of the mean.}
  \label{fig:tradeoff}
\end{figure*}

\begin{figure*}[h!]
  \centering
  \includegraphics[width=\textwidth]{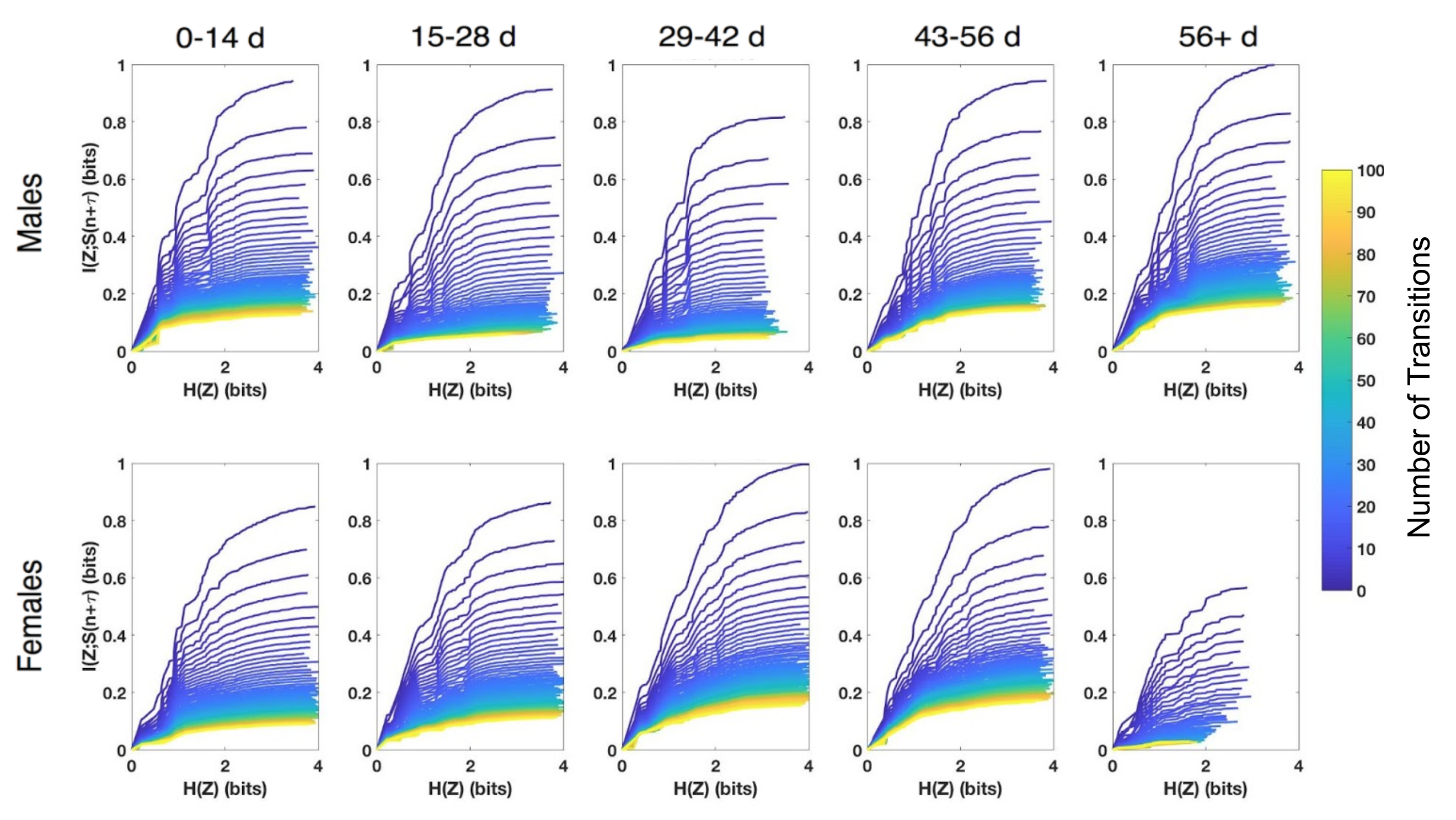}
  \caption{Trade-off curves computed from the deterministic information bottleneck for each sex and age group.}
  \label{fig:tradeoff2}
\end{figure*}

%\bibliography{mylib}

\end{document}